# A Differential Evaluation Markov Chain Monte Carlo algorithm for Bayesian Model Updating

M. Sherri [a], I. Boulkaibet [b], T. Marwala [b], M. I. Friswell [c],

[a] Department of Mechanical Engineering Science, University of Johannesburg, PO Box 524, Auckland Park 2006, South Africa.
[b] Institute of Intelligent Systems, University of Johannesburg, PO Box 524, Auckland Park 2006, South Africa.
[c] College of Engineering, Swansea University, Bay Campus, Swansea SA1 8EN, United Kingdom.

**Abstract**:

The use of the Bayesian tools in system identification and model updating paradigms has been increased in the last ten years. Usually, the Bayesian techniques can be implemented to incorporate the uncertainties associated with measurements as well as the prediction made by the finite element model (FEM) into the FEM updating procedure. In this case, the posterior distribution function describes the uncertainty in the FE model prediction and the experimental data. Due to the complexity of the modeled systems, the analytical solution for the posterior distribution function may not exist. This leads to the use of numerical methods, such as Markov Chain Monte Carlo techniques, to obtain approximate solutions for the posterior distribution function. In this paper, a Differential Evaluation Markov Chain Monte Carlo (DE-MC) method is used to approximate the posterior function and update FEMs. The main idea of the DE-MC approach is to combine the Differential Evolution, which is an effective global optimization algorithm over real parameter space, with Markov Chain Monte Carlo (MCMC) techniques to generate samples from the posterior distribution function. In this paper, the DE-MC method is discussed in detail while the performance and the accuracy of this algorithm are investigated by updating two structural examples.

1. Introduction

During the last thirty years, the application of the finite element method (FEM) [1-3] has exponentially increased where this numerical technique has become one of the most popular engineering tools in systems modelling and prediction. In the domain of structural dynamics, the FEM tools are widely applied to model complex systems where this technique can produce results with high accuracy, especially when the modelled system is simple. However, the results attained by the FEM can be relatively inaccurate and the mismatches between the FEM results and the results attained from experimental studies are relatively significant. This is due to the errors associated with the modelling process as well as the complexity of modelled structure, which may reduce the accuracy of the modelling process. Consequently, the model obtained by an FEM needs to be updated to reduce the errors between the experimental and modelled outputs. The procedure of minimizing the differences between the numerical results and the measured data is known as the finite element model updating (FEMU) [4, 5], where the FEMU methods can be divided into two main classes. In the first class, which is the direct methods, we equate the experimental data directly to the FEM outputs resulting in a procedure that constrains the updating to the FE system matrices (mass, stiffness) only. This kind of approach may produce non-realistic results where the resulting updating parameters may not have physical meaning. In the second class, which is also known as the iterative (or indirect) approaches, the FEM outputs are not directly equated to the experimental data, but instead, an objective function is introduced and iteratively minimised to reduce the errors between the analytical and experiential results. Thus, we vary the system matrices and the model output during the minimisation process, and realistic results are often expected at the end of the updating process.

Generally, several sources of uncertainty are associated with the modelling process, such as the mathematical simplifications made during the modelling, where this kind of uncertainty may affect the accuracy of the modelling process. Moreover, the noise that contaminates the experimental results may also have a significant impact on the updating process. To deal with such uncertainty problems, another class of methods called the uncertainty quantification methods accomplishes the updating process. The most common uncertainty quantification method is known as the Bayesian approach in which the unknown parameters and their uncertainty are identified by defining each unknown parameter with a probability density distribution (PDF). Recently, the use of the Bayesian methodology has massively increased in the domain of system identification and uncertainty quantification. In this approach, the uncertainties associated with the modelled structure are expressed in terms of probability distributions where the unknown parameters are defined as a random vector with a multi-variable probability density function, and the resulting function is known as the posterior PDF. Solving the posterior PDF helps in identifying the unknown parameters and their uncertainties. Unfortunately, the posterior PDF cannot be solved in an analytical way for sufficiently complex problems which is the case for the FEMU problems since the search space is usually nonlinear and high dimensional. In this case, sampling techniques are employed to identify these uncertain parameters. The most recognised sampling methods are these related to Markov chain Monte Carlo (MCMC) methods.

Generally, the MCMC methods are very useful tools that can efficiently cope with large search spaces and generate samples from complex distributions. These methods draw samples with an element of randomness while being guided by the values of the posterior distribution function. Then, the drawn samples are accepted or rejected according to the Metropolis criterion. Unfortunately, the updated models, with relatively large complexities, may have multiple optimal (or near optimal) simple MCMC algorithms cannot easily identify solution, and this. In this paper, another version of the MCMC algorithms, known as the Differential Evolution Markov Chain (DE-MC) [6, 7] algorithm, is used to update FEMs of structural systems. The DE-MC algorithm combines the abilities of the differential evaluation algorithm [8, 9], which is one of the genetic algorithms for global optimization, with the Metropolis-Hastings algorithm. In this algorithm, multiple chains are run in parallel, and the exploration and exploitation of the search space in the current chain are achieved by the difference of two randomly selected chains, multiplied by the value of the difference with a preselected factor and then the result is added to the value of the current chain. The value of the current chain is then accepted or rejected according to the Metropolis criterion. In this paper, the efficiency, reliability and the limitations of the DE-MC algorithm are investigated when the Bayesian approach is applied for FEMU. This paper is organized as follows: in the next section, the Bayesian formulations are introduced. Section 3 describes the DE-MC algorithm while section 4 presents the results when a simple mass-spring structure is updated. Section 5 presents the updating results of an unsymmetrical H-shaped Structure. The paper is concluded in section 6.

2. **Bayesian formulations**

In this paper, the Bayesian approach is adopted to compute the posterior distribution function in order to update the FEMs. The posterior function can be represented by Bayes rule [10-14]:

$$P(\boldsymbol{\theta}|\mathcal{D},\mathcal{M}) \propto P(\mathcal{D}|\boldsymbol{\theta},\mathcal{M}) \, P(\boldsymbol{\theta}|\mathcal{M}) \qquad (2.1)$$

where $\mathcal{M}$ describes the model class for the target system where each model class $\mathcal{M}$ is defined by certain updating parameters $\boldsymbol{\theta} \in \boldsymbol{\Theta} \subset \mathcal{R}^d$. The experimental data $\mathcal{D}$ of the structural system, which is represented by the natural frequencies $f_i^m$ and mode shapes $\boldsymbol{\phi}_i^m$, are used to improve the FEM results. $P(\boldsymbol{\theta}|\mathcal{M})$ is the prior probability distribution function (PDF) that represents the initial knowledge of the uncertain parameters given a specific model $\mathcal{M}$, and in the absence of the measured data $\mathcal{D}$. The function $P(\mathcal{D}|\boldsymbol{\theta},\mathcal{M})$ is known as the likelihood function and represents the difference between the experimental data and the FEM results. Finally, the probability distribution function $P(\boldsymbol{\theta}|\mathcal{D},\mathcal{M})$ is the posterior function of the unknown parameters given a model class $\mathcal{M}$ and the measured data $\mathcal{D}$. The model class $\mathcal{M}$ is used only when several classes are investigated for both model updating and model selection. In this paper, only one model class is considered, and therefore, the term $\mathcal{M}$ is omitted in order to simplify the Bayesian formulations.

In this paper, the likelihood function is given by:

$$P(\mathcal{D}|\boldsymbol{\theta}) = \frac{1}{\left(\frac{2\pi}{\beta_c}\right)^{N_m/2} \prod_{i=1}^{N_m} f_i^m} \exp\left(-\frac{\beta_c}{2} \sum_i^{N_m} \left(\frac{f_i^m - f_i}{f_i^m}\right)^2\right) \tag{2.2}$$

where $N_m$ is the number of measured modes, $\beta_c$ is a constant, $f_i^m$ and $f_i$ are the $i$th analytical and measured natural frequencies. The initial knowledge of the updating parameters $\boldsymbol{\theta}$, which is defined by a prior PDF, is given by the following Gaussian distribution:

$$P(\boldsymbol{\theta}) = \frac{1}{(2\pi)^{Q/2} \prod_{i=1}^{Q} \frac{1}{\sqrt{\alpha_i}}} \exp\left(-\sum_i^Q \frac{\alpha_i}{2} \|\theta^i - \theta_0^i\|^2\right) = \frac{1}{(2\pi)^{Q/2} \prod_{i=1}^{Q} \frac{1}{\sqrt{\alpha_i}}} \exp\left(-\frac{1}{2}(\boldsymbol{\theta} - \boldsymbol{\theta}_0)^T \Sigma^{-1} (\boldsymbol{\theta} - \boldsymbol{\theta}_0)\right) \tag{2.3}$$

where $Q$ is the number of the uncertain parameters, $\boldsymbol{\theta}_0$ represents the mean value of the updating parameters, $\alpha_i, i = 1, \ldots, Q$ are the coefficients of the updating parameters and the Euclidean norm is given by the notation: $\|*\|$.

After substituting Eqs. (2.2) and (2.3) into the Bayesian inference defined by Eq. (2.1), the posterior $P(\boldsymbol{\theta}|\mathcal{D})$ of the unknown parameters $\boldsymbol{\theta}$ given the experimental data $\mathcal{D}$ is characterized by:

$$P(\boldsymbol{\theta}|\mathcal{D}) \propto \frac{1}{Z_s(\alpha, \beta_c)} \exp\left(-\frac{\beta_c}{2} \sum_i^{N_m} \left(\frac{f_i^m - f_i}{f_i^m}\right)^2 - \sum_i^Q \frac{\alpha_i}{2} \|\theta^i - \theta_0^i\|^2\right) \tag{2.4}$$

where

$$Z_s(\alpha, \beta_c) = \left(\frac{2\pi}{\beta_c}\right)^{N_m/2} \prod_{i=1}^{N_m} f_i^m (2\pi)^{Q/2} \prod_{i=1}^{Q} \frac{1}{\sqrt{\alpha_i}} \tag{2.5}$$

Generally, the complexity of the posterior PDF, which depends on the modal parameters of the analytical model, is related to the complexity of the analytical model, and for certain relatively complex structural models the analytical results for the posterior distribution are difficult to obtain due to the high dimensionality of the search space. In this case, sampling techniques [5, 10, 11, 13, 14] are the only practical approaches in order to approximate the posterior PDF. The main idea of sampling techniques is to generate a $N_s$ sequence of vectors $\{\boldsymbol{\theta}_1, \boldsymbol{\theta}_2, \ldots, \boldsymbol{\theta}_{N_s}\}$ and use these samples to approximate the future response of the unknown parameters at different time instances. The most recognized sampling techniques are Markov Chain Monte Carlo (MCMC) methods [5, 13-18]. In this paper, the combination of one of the basic MCMC algorithms, known as the Metropolis-Hasting algorithm, with one of the genetic algorithms, known as differential evolution (DE), is used to generate samples from the posterior PDF in order to update structural models.

### 3. The Differential Evolution Markov Chain Monte Carlo (DE-MC) method

In this paper, the DE and MCMC methods, which are extremely popular methods in several scientific domains, are combined to improve the convergence of the sampling procedure. In this approach, multiple chains are run in parallel in order to improve the accuracy of the updating parameters, while these chains learn from each other instead of running all the chains independently. This may improve the efficiency of the searching procedure and avoid sampling in the vicinity of a local minimum. The new chains are then accepted or rejected according to the Metropolis-Hastings criterion.

The Metropolis-Hastings (M-H) [18, 19, 20] algorithm is one of the common MCMC methods that can be used to draw samples from multivariate probability distributions. To sample from the posterior PDF $P(\theta|D)$, where $\boldsymbol{\theta} = \{\theta_1, \theta_2, \ldots, \theta_d\}$ is a $d$-dimensional parameters vector, a proposal density distribution $q(\boldsymbol{\theta}|\boldsymbol{\theta}_{t-1})$ is used to generate a proposed random vector $\boldsymbol{\theta}^*$ given the value at the previous accepted vector $\boldsymbol{\theta}_{t-1}$ at the iteration $t-1$ of the algorithm. Next, the Metropolis criterion is used to accept or reject the proposed sample $\boldsymbol{\theta}^*$ as follows:

$$\alpha(\boldsymbol{\theta}^*|\boldsymbol{\theta}_{t-1}) = \min\left\{1, \frac{P(\boldsymbol{\theta}^*|D)\, q(\boldsymbol{\theta}_{t-1}|\boldsymbol{\theta}^*)}{P(\boldsymbol{\theta}_{t-1}|D)q(\boldsymbol{\theta}^*|\boldsymbol{\theta}_{t-1})}\right\} \qquad (3.1)$$

On the other hand, the Differential Evolution (DE) [8] is a very effective genetic algorithm in solving various real-world global optimization problems. As one of the genetic algorithms, the DE algorithm begins by randomly initialising the population within certain search area, and then these initial values are evolved over the generations in order to find the global minimum. This can be achieved using genetic operators such as: mutation, selection, and crossover.

By integrating the Metropolis-Hastings criterion within the search abilities of the DE algorithm, the resulted MCMC method can be more efficient in determining where other chains can be employed to create the new candidates for the current chain. In the DE-MC algorithm, the new value of the chain is obtained by a simple mutation operation where the difference between two randomly selected chains (different from the current chain) are added to the current chain. Thus, the proposal for each chain depends on a weighted combination of other chains which can be easily defined as [6, 7]:

$$\boldsymbol{\theta}^* = \boldsymbol{\theta}_i + \gamma(\boldsymbol{\theta}_a - \boldsymbol{\theta}_b) + \boldsymbol{\varepsilon} \qquad (3.2)$$

where $\boldsymbol{\theta}^*$ represents the new proposed vector, $\boldsymbol{\theta}_i$ is the current state of the $i$-th chain, $\boldsymbol{\theta}_a$ and $\boldsymbol{\theta}_b$ are randomly selected chains, $\gamma$ is a tuning factor that always take a positive value and can be set to vary between $[0.4, 1]$. Note that the vectors: $\boldsymbol{\theta}_i \neq \boldsymbol{\theta}_a \neq \boldsymbol{\theta}_b$. Finally, the noise $\boldsymbol{\varepsilon}$, which is defined as a Gaussian distribution $\boldsymbol{\varepsilon} \sim N_p(\boldsymbol{0}, \boldsymbol{\sigma}^2)$ with a very small variance vector $\boldsymbol{\sigma}^2$, is added to the proposed vector to avoid degeneracy problems. The factor $\gamma$ can be seen as the magnitude that controls the jumping distribution. The main idea of the DE-MC algorithm can be illustrated in Figure 3.1b.

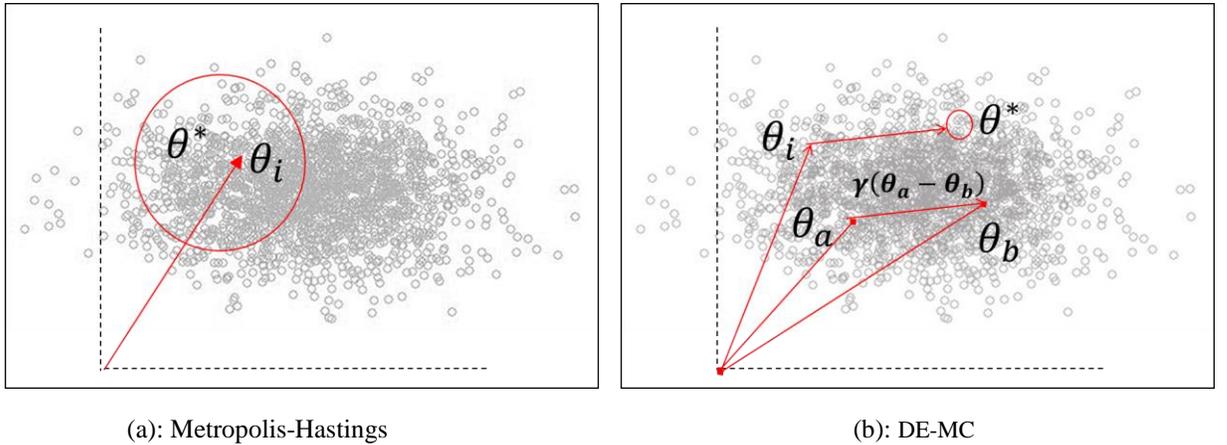

(a): Metropolis-Hastings  (b): DE-MC

Figure 3.1: Proposed vector generation in the M-H and DE-MC methods

Figure 3.1 explains the way to generate proposed vectors for the M-H method (Figure 3.1a) and for the DE-MC method (Figure 3.1b). As illustrated, the difference vector between the two randomly selected chains $\boldsymbol{\theta}_a$ and $\boldsymbol{\theta}_b$ represents the direction of the new proposed vector, where this difference is multiplied by the factor $\gamma$ to define the moving distance. The moving distance is then added to the current chain $\boldsymbol{\theta}_i$ to create the proposed vector. Note that, the DE-MC method has only one tuning factor $\gamma$ in comparing to other versions of evolutionary MCMC methods. Finally, the new proposal $\boldsymbol{\theta}^*$ of the $i$-th chain is accepted or rejected according to the Metropolis criterion which is given as:

$$r = \min\left\{1, \frac{P(\boldsymbol{\theta}^*|D)}{P(\boldsymbol{\theta}_i|D)}\right\} \tag{3.3}$$

The steps to update FEMs using the DE-MC algorithm are summarized as follows:

1- Initialize the population $\boldsymbol{\theta}_{i,o}$, $i \in \{1,2,\ldots,N\}$.
2- Set the tuning factor $\gamma$. In this paper, $\gamma = 2.38/\sqrt{2d}$ and $d$ is the dimension of the updating parameters.
3- Calculate the Posterior PDF for all chains.
4- For all chains $i \in \{1,2,\ldots,N\}$:
   4.1 Sample uniformly two random vectors $\boldsymbol{\theta}_a, \boldsymbol{\theta}_b$ where $\boldsymbol{\theta}_a \neq \boldsymbol{\theta}_b \neq \boldsymbol{\theta}_i$.
   4.2 Sample the random value $\boldsymbol{\varepsilon}$ with small variance $\boldsymbol{\varepsilon} \sim N_p(\boldsymbol{0}, \boldsymbol{\sigma}^2)$.
   4.3 Calculate the proposed vector $\boldsymbol{\theta}^* = \boldsymbol{\theta}_i + \gamma(\boldsymbol{\theta}_a - \boldsymbol{\theta}_b) + \boldsymbol{\varepsilon}$.
   4.4 Calculate the Posterior PDF for the vector $\boldsymbol{\theta}^*$.
   4.5 Calculate the Metropolis ratio $r = \min\left\{1, \frac{P(\boldsymbol{\theta}^*|D)}{P(\boldsymbol{\theta}_i|D)}\right\}$
   4.6 Accept the proposed vector $\boldsymbol{\theta}_i \leftarrow \boldsymbol{\theta}^*$ with probability $min(1,r)$, otherwise $\boldsymbol{\theta}_i$ is unchanged.

5- Repeat the steps 4.1 to 4.6 until the number of samples required is achieved.

In next two sections, the DE-MC performance is highlighted when two structural examples are updated.

**4. Application 1: Simple Mass-Spring system**

In this section, a five degrees of freedom mass-spring linear system, as presented in Figure 4.1, is updated using the DE-MC algorithm.

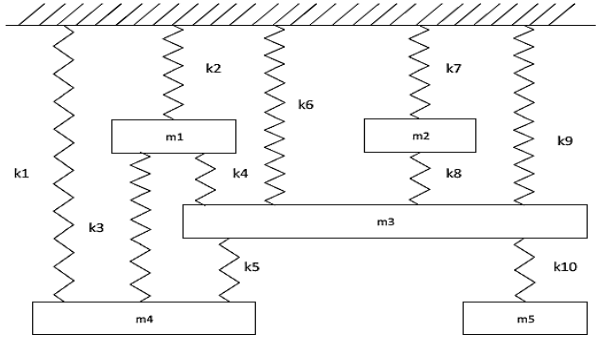

Figure 4.1: The five degrees of freedom mass-spring system

The system contains 5 masses connected to each other using 10 springs (see Figure 4.1). The deterministic values of the masses are: $m_1 = 2.7$ kg, $m_2 = 1.7$ kg, $m_3 = 6.1$ kg, $m_4 = 5.3$ kg and $m_5 = 2.9$ kg. The stiffness of the springs are: $k_3 = 3200$ N/m, $k_5 = 1840$ N/m, $k_7 = 2200$ N/m, $k_9 = 2800$ N/m and $k_{10} = 2000$ N/m. The spring stiffnesses $k_1, k_2, k_4, k_6$, and $k_8$ are considered as the uncertain parameters where the updating vector is: $\boldsymbol{\theta} = \{\theta_1, \theta_2, \theta_3, \theta_4, \theta_5\} = \{k_1, k_2, k_4, k_6, k_8\}$.

Since the DE-MC method is used for the updating procedure, the population used by the algorithm is selected to be $N = 10$. The updating vectors are bounded by $\boldsymbol{\theta}_{max}$ and $\boldsymbol{\theta}_{min}$ which are set to $\{4800, 2600, 2670, 3400, 2750\}$ and $\{3200, 1800, 1600, 1800, 2050\}$, respectively. The tuning factor is set to $\gamma = 2.38/\sqrt{2d}$ while $d = 5$, the initial vector of $\boldsymbol{\theta}$ is set to $\boldsymbol{\theta}_0 = \{4600, 2580, 1680, 3100, 2350\}$ and the number of generations (number of samples) is set to $N_s = 10000$. The obtained samples are illustrated in Figure 4.2 while the updating parameters, as well as the initial and updated natural frequencies, are shown in Tables 4.1 and 4.2, respectively.

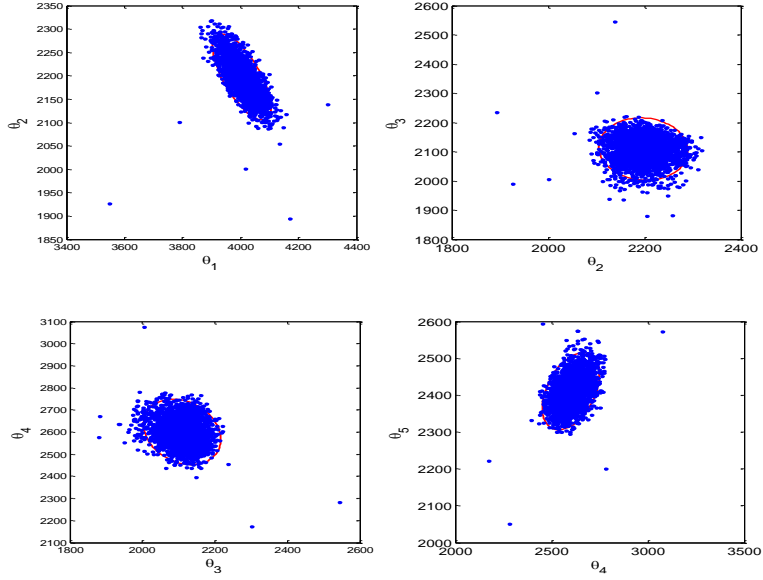

Figure 4.2: The scatter plots of the samples using the DE-MC algorithm

Figure 4.2 shows the scatter plots of the uncertain parameters using the DE-MC algorithm. The confidence ellipses (error ellipse) of the samples are also shown in the same figures (in red) where these ellipses visualize the regions that contain 95% of the obtained samples. As expected, the figure shows that the DE-MC algorithm has found the high probability area after only few iterations. Table 4.1 contains the initial values, the nominal values and the updated values of the uncertain parameters. The coefficient of variation (c.o.v) values, which are estimated by dividing the standard deviations $\sigma_i$ by the updated vectors $\boldsymbol{\theta}_i$ (or $\boldsymbol{\mu}_i$), are also presented in Table 4.1 and used to measure the errors in the updating. It is clear that the obtained values of the c.o.v when the DE-MC algorithm is used to update the structure are small and less than 2.5% which means that the DE-MC algorithm performed well and was able to identify the areas with high probability. This also can be verified from the same table where the updating parameters are close the nominal values.

Table 4.1: The updating parameters using DE-MC technique

| | Unknown parameters (N/m) | | | | | |
|---|---|---|---|---|---|---|
| | Initial | Nominal values | Error (%) | DE-MC ($\mu_i$) | Error (%) | $\frac{\sigma_i}{\mu_i}$ c.o.v %) |
| $\theta_1$ | 4600 | 4010 | 14.71 | 4004.4 | 0.14 | 1.03 |
| $\theta_2$ | 2580 | 2210 | 16.74 | 2197.6 | 0.56 | 1.71 |
| $\theta_3$ | 1680 | 2130 | 21.13 | 2109.4 | 0.97 | 2.07 |
| $\theta_4$ | 3100 | 2595 | 19.46 | 2600.9 | 0.23 | 2.32 |
| $\theta_4$ | 2350 | 2398 | 02.00 | 2410.4 | 0.52 | 1.77 |

Table 4.2 contains the initial, nominal and updated natural frequencies. Furthermore, the absolute errors, which are estimated by $\frac{|f_i^m - f_i|}{f_i^m}$, the total average error ($TAE$), which is computed by $TAE = \frac{1}{N_m}\sum_{i=1}^{N_m} \frac{|f_i^m - f_i|}{f_i^m}$, $N_m = 5$, and the c.o.v values are also displayed. Obviously, the updated frequencies obtained by the DE-MC are better than the initial frequencies, and almost equal to the nominal frequencies.

Table 4.2: The updated natural frequencies and the errors obtained using the DE-MC

| Modes | Nominal Frequency (Hz) | Initial Frequency (Hz) | Error (%) | Frequency DE-MC (Hz) | c.o.v (%) | Error (%) |
|---|---|---|---|---|---|---|
| 1 | 3.507 | 3.577 | 1.97 | 3.507 | 0.118 | 0.00 |
| 2 | 5.149 | 5.371 | 4.30 | 5.149 | 0.126 | 0.00 |
| 3 | 7.083 | 7.239 | 2.21 | 7.082 | 0.119 | 0.02 |
| 4 | 8.892 | 9.030 | 1.56 | 8.894 | 0.140 | 0.03 |
| 5 | 9.426 | 9.412 | 0.16 | 9.426 | 0.117 | 0.00 |
| TAE | ——— | ——— | 1.98 | ——— | —— | **0.012** |

The total average error of the FEM output was reduced from 1.98% to 0.012%. On the other hand, the values of the c.o.v for all updated frequencies are smaller than 0.15% which indicates that the DE-MC technique efficiently updated the structural system. Figure 4.3 shows the evaluation of the total average error at each iteration. The $TAE$ in Figure 4.3 is obtained as follows: first, the mean value of the samples at each iteration is computed as $\hat{\theta} = E(\theta) \cong \frac{1}{N_s}\sum_{j=1}^{i} \theta^i$ where $i$ is the current iteration. Next, the mean value is used to compute the analytical frequencies of the FEM, and then the total average error is calculated as: $TAE(i) = \frac{1}{N_m}\sum_{j=1}^{N_m} \frac{|f_j^m - f_j|}{f_j^m}$. As a result, it is clear that the DE-MC algorithm converges efficiently after the first 2000 iterations.

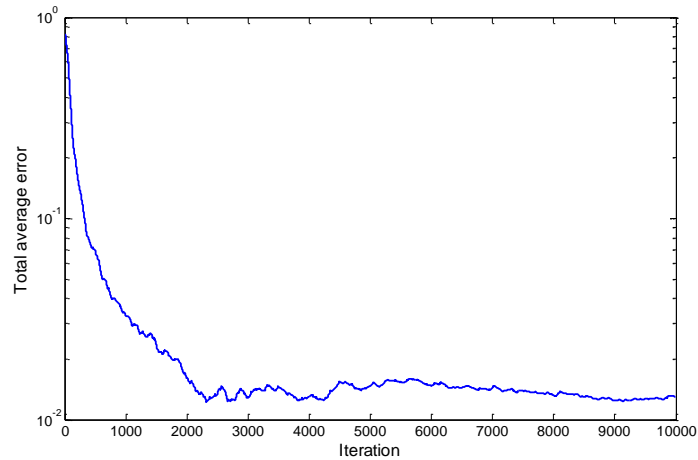

Figure 4.3: The evaluation of the TAE using the DE-MC method

Figure 4.4 illustrates the correlation between the updating parameters where all parameters are correlated (the values are different from zero). Moreover, the majority of these parameters are weakly correlated (small values <0.3) except the pairs ($\theta_1, \theta_2$) and ($\theta_4, \theta_5$) which are highly correlated (values >0.7).

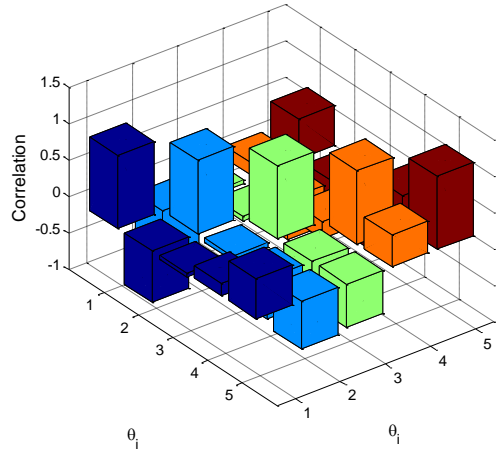

Figure 4.4: The correlation between the updating parameters.

In the next section, the DE-MC method is used to update an unsymmetrical H-shaped aluminum structure with real experimental data.

## 5. Application 2: The FEMU of the unsymmetrical H-Shaped structural system.

In this section, the performance of the DE-MC algorithm is examined by updatintg an unsymmetrical H-shaped aluminum structure with real measured data. The FEM model of the H-shaped structure is presented in figure (5.1) where the structure is divided into 12 elements, and each element is modelled as an Euler-Bernoulli beam. The location displayed by a double arrow at the middle beam indicates the position of excitation which is produced by an electromagnetic shaker. An accelerometer was used to measure the set of frequency-response functions. The initial analytical natural frequencies are 53.9, 117.3, 208.4, 254.0 and 445.0 Hz. In this example, the moments of inertia $I_{xx}$ and the cross-sectional areas $A_{xx}$ of the left, middle and right subsections of the H-shaped beam are selected to be updated in order to improve the analytical natural frequencies. Thus, the updating parameters are: $\boldsymbol{\theta} = \{I_{x1}, I_{x2}, I_{x3}, A_{x1}, A_{x2}, A_{x3,}\}$.

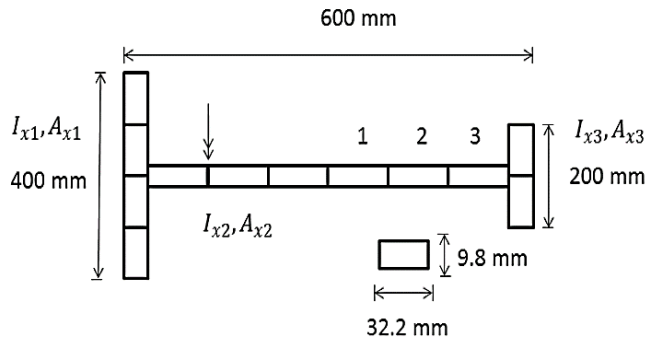

Figure 5.1: The Unsymmetrical H-Shaped Structure

The rest of the H-shaped structure parameters are given as follows: The Young's modulus is set to $7.2 \times 10^{10}$ N/m² and the density is set to 2785 kg/m³. The updating parameters $\boldsymbol{\theta}$ are bounded by maximum and minimum vectors given by: $[4.73 \times 10^{-8}, 4.73 \times 10^{-8}, 4.73 \times 10^{-8}, 5.16 \times 10^{-4}, 5.16 \times 10^{-4}, 5.16 \times 10^{-4}]$ and $[0.73 \times 10^{-8}, 0.73 \times 10^{-8}, 0.73 \times 10^{-8}, 1.16 \times 10^{-4}, 1.16 \times 10^{-4}, 1.16 \times 10^{-4}]$, respectively. These boundaries are used to ensure that the updating parameters are physically realistic. The number of samples is set to $N_s = 5000$, the factor $\beta_c$ of the likelihood function is set equal to 10, the coefficients $\alpha_i$ of the prior PDF are set to $\frac{1}{\sigma_i^2}$ where $\sigma_i^2$ is the variance of the $i$th uncertain parameters, and $\sigma = [5 \times 10^{-8}, 5 \times 10^{-8}, 5 \times 10^{-8}, 5 \times 10^{-4}, 5 \times 10^{-4}, 5 \times 10^{-4}]$.

Figure 5.2 illustrates the scatter plots of the updating parameters. The confidence ellipse that contains 95% of samples are also included in the figure. The updating parameters were normalized by dividing the parameters by $k = [10^{-8}, 10^{-8}, 10^{-8}, 10^{-4}, 10^{-4}, 10^{-4}]$. As expected, the DE-MC algorithm was able to find the area with high probably after a few iterations. The rest of the updating parameters are shown in Table 5.1 as well as the initial values, the c.o.v values and the updating parameters obtained by the M-H algorithm.

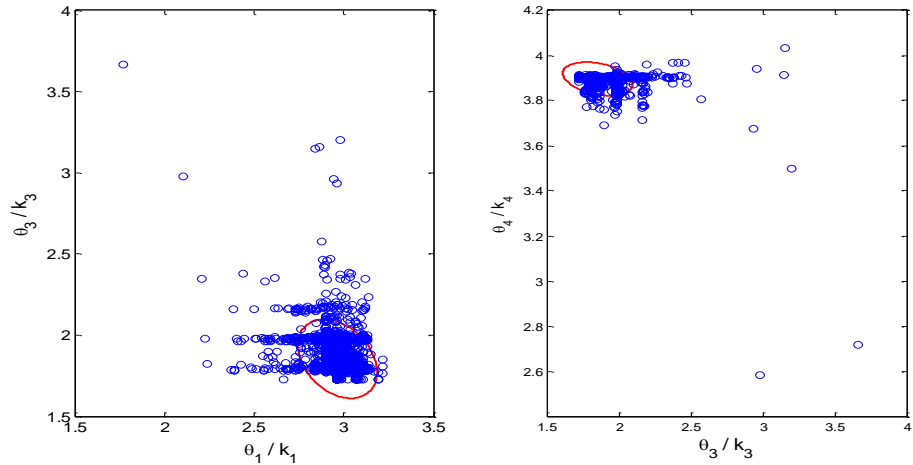

Figure 5.2: The scatter plots of the samples using the DE-MC algorithm

Table 5.1: The initial parameters, the c.o.v values and the updating parameters using the DE-MC and M-H algorithms

|  | Initial | DE-MC ($\mu_i$) | $\frac{\sigma_i}{\mu_i}$ (%) | M-H ($\mu_i$) | $\frac{\sigma_i}{\mu_i}$ (%) |
|---|---|---|---|---|---|
| $\theta_1$ | $2.7265 \times 10^{-8}$ | $2.8965 \times 10^{-8}$ | 5.76 | $2.31 \times 10^{-8}$ | 22.59 |
| $\theta_2$ | $2.7265 \times 10^{-8}$ | $2.9739 \times 10^{-8}$ | 1.91 | $2.68 \times 10^{-8}$ | 15.25 |
| $\theta_3$ | $2.7265 \times 10^{-8}$ | $1.7676 \times 10^{-8}$ | 1.67 | $2.17 \times 10^{-8}$ | 13.96 |
| $\theta_4$ | $3.1556 \times 10^{-4}$ | $3.8966 \times 10^{-4}$ | 0.65 | $2.85 \times 10^{-4}$ | 14.36 |
| $\theta_5$ | $3.1556 \times 10^{-4}$ | $2.1584 \times 10^{-4}$ | 2.93 | $2.83 \times 10^{-4}$ | 14.36 |
| $\theta_6$ | $3.1556 \times 10^{-4}$ | $2.9553 \times 10^{-4}$ | 0.026 | $2.77 \times 10^{-4}$ | 13.08 |

The results in Table 5.1 indicate that the updating parameters obtained by the DE-MC and M-H algorithms are different from the initial values which means that the uncertain parameters have been successfully updated. Furthermore, the c.o.v values of the updating parameters obtained by the DE-MC algorithm are relatively small (<2.5%) with verifies that the algorithm was able to identify the areas with high probability in a reasonable amount of time; however, the c.o.v obtained by the M-H algorithm are relatively high (>13.08%) which means that the M-H algorithm does not have the efficiency of the DE-MC algorithm.

Table 5.2 illustrates the updating frequencies using the DE-MC and M-H algorithms, the errors and the c.o.v values. As expected, the analytical frequencies obtained by the DE-MC algorithm are better than the initial frequencies as well as the frequencies obtained by the M-H algorithm. The DE-MC method has improved all natural frequencies and reduced the total average error (TAE) from 5.37% to 1.53%. Also, the c.o.v values obtained by the DE-MC method are relatively small (<0.65%).

Table 5.2: Natural frequencies, c.o.v values and errors when DE-MC and M-H techniques are used for FEMU

| Modes | Measured Frequency (Hz) | Initial Frequency (Hz) | Error (%) | Frequency DE-MC (Hz) | c.o.v (%) | Error (%) | Frequency M-H (Hz) | c.o.v (%) | Error (%) |
|---|---|---|---|---|---|---|---|---|---|
| 1 | 53.90 | 51.04 | 5.31 | 52.56 | 0.30 | 2.49 | 53.92 | 3.96 | 0.04 |
| 2 | 117.30 | 115.79 | 1.29 | 119.42 | 0.35 | 1.81 | 122.05 | 4.28 | 4.05 |
| 3 | 208.40 | 199.88 | 4.09 | 210.46 | 0.54 | 0.99 | 210.93 | 4.95 | 1.22 |
| 4 | 254.00 | 245.76 | 3.25 | 253.37 | 0.41 | 0.25 | 258.94 | 4.81 | 1.94 |
| 5 | 445.00 | 387.53 | 12.92 | 435.71 | 0.63 | 2.09 | 410.33 | 4.74 | 7.79 |
| TAE | _____ | _____ | 5.37 | _____ | _____ | 1.53 | _____ | _____ | 3.01 |

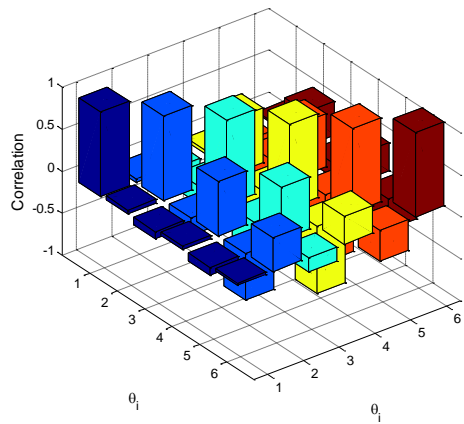

Figure 5.3: The correlation between the updating parameters

Figure 5.3 shows the correlation of the updating parameters using the DE-MC parameters where the majority of these parameters are weakly correlated except the pairs ($\theta_2, \theta_5$) and ($\theta_4, \theta_5$), where the correlation between these pairs are relatively high (<0.7%).

The evaluation of total average error after each accepted (or rejected) sample is illustrated in Figure 5.4. The result indicates that the DE-MC has a fast convergence rate and was able converge after 500 iterations.

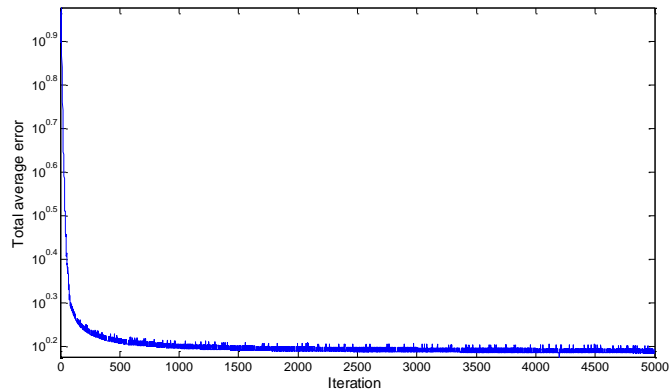

Figure 5.4: The evaluation of the TAE using the DE-MC method

### 6. Conclusion

In this paper, the Differential Evolution Markov Chain (DE-MC) algorithm is used to approximate the Bayesian formulations in order to perform a finite element model updating procedure. In the DE-MC method, multi-chains are run in parallel which allows the chains to learn from each other in order to improve the sampling process where the jumping step depends on the difference between randomly selected chains. This method is investigated by updating two structural system: the first one is a five DOF mass-spring linear system and the second one is the unsymmetrical H-shaped aluminum structure. In the first case, the total average error was reduced from 1.98% to 0.012%, while in the second case, the FEM updating of the unsymmetrical H-shaped structure, the total average error was reduced from 5.37% to 1.53%. Also the DE-MC algorithm appeared to have better results than the M-H algorithm when the unsymmetrical H-shaped structure is updated. In further work, the DE-MC algorithm will be modified and improved to include several steps such as the cross over and the exchange between the parallel chains. These changes may further improve the sampling procedure.